\documentclass{aa}  

\usepackage{graphicx}

\usepackage{txfonts}
\usepackage[font=small]{caption}
\usepackage{geometry}
\usepackage{array}
\usepackage{makecell}
\usepackage{dblfloatfix}
\usepackage{float}
\usepackage{caption}
\usepackage{multirow}
\usepackage{subcaption}
\usepackage{hyperref}
\hypersetup{
    colorlinks=true,
    linkcolor=red,
    urlcolor=cyan,
    citecolor=blue
    }
\usepackage[capitalize]{cleveref}
\usepackage{tablefootnote}
\usepackage{longtable}
\usepackage{longtable}
\usepackage{footmisc}
\usepackage{gensymb}
\usepackage{xcolor}
\usepackage{booktabs}
\newcolumntype{P}[1]{>{\centering\arraybackslash}p{#1}}

\begin{document} 

   \title{\textit{TUVO-21acq}: a new cataclysmic variable discovered through a UV outburst\thanks{A reproduction package for this paper will be made available at \url{https://doi.org/10.5281/zenodo.6483027}.}}

   \author{D. Modiano\inst{1}, R. Wijnands\inst{1}, D.A.H Buckley\inst{2,3,4}, M. Gromadzki\inst{5}, S. Verberne\inst{1,6}, M. van Etten\inst{1}}
   \institute{Anton Pannekoek Institute for Astronomy, University of Amsterdam, Postbus 94249, 1090 GE Amsterdam, The Netherlands\\
   \email{d.modiano@uva.nl} \and
   South African Astronomical Observatory, PO Box 9, Observatory Road, Observatory 7935, Cape Town, South Africa  \and
   Department of Astronomy, University of Cape Town, Private Bag X3, Rondebosch 7701, South Africa  \and
   Department of Physics, University of the Free State, PO Box 339, Bloemfontein 9300, South Africa  \and
   Astronomical Observatory, University of Warsaw, Al. Ujazdowskie 4, 00-478 Warszawa, Poland \and
   Leiden Observatory, Leiden University, PO Box 9513, 2300 RA, Leiden, The Netherlands}
 
  \abstract
  {Outbursts from cataclysmic variables, such as dwarf novae (DNe), are prevalent throughout the galaxy and are known to emit strongly in the ultraviolet (UV). However, the UV emission of DN outbursts has not been studied extensively compared with the optical. Characterising in detail the physical processes responsible for outburst behaviour requires further UV data, since the UV probes the inner regions of the accretion disk. Here we report, as part of our recently launched Transient UV Objects Project (TUVO), the discovery of TUVO-21acq, a new transient which we detected in the UV using data from the Ultraviolet Optical Telescope (UVOT) aboard \textit{Swift}. We detected two separate outbursts and used the UVOT data to constrain source properties, focusing on the amplitudes and timescales of the outbursts. We found that during the first outburst the source increased in brightness by at least 4.1, 2.4, and 3.5 magnitudes and during the second outburst by 4.4, 3.4, and 3.6 magnitudes in the UVW1, UVM2, and UVW2 bands, respectively. The outburst durations were in the range 6-21 days and 11-46 days, and we determined an upper limit for the recurrence time of 316 days. To further characterise the source, we obtained an optical spectrum during quiescence with SALT. The spectrum exhibited hydrogen Balmer series and helium emission lines, and a flat overall spectral shape. We confirm the nature of the source as an accreting white dwarf which underwent DN outburst based on photometric and spectroscopic properties. This study serves as a proof-of-concept for the TUVO project strategy, demonstrating that it has the capability of discovering and classifying new, interesting UV transients. We also discuss the implications of our findings for our understanding of the physics underlying DN outbursts, in particular with respect to the UV emission. We examine the need for simultaneous UV and optical observations during the onset of DN outbursts in order to help answer remaining questions, for example in the characteristics and implications of the UV delay.}

   
    \titlerunning{UV outburst of a new CV}
    \authorrunning{Modiano et al.}
   
   \maketitle

\section{Introduction}  \label{introduction}

Cataclysmic variables (CVs) are semi-detached binary systems consisting of a white dwarf (WD) primary that is accreting mass from a secondary low-mass companion (see \citealp{warner_1995}). In systems in which the WD has no magnetic field (or a very weak magnetic field), accretion occurs via an accretion disk that forms around the WD. If the WD has a moderate or strong magnetic field (at least 10$^{7}$G; \citealp{Zhilkin_2012,Inight_2021_2}), the disk will be truncated or entirely prevented from forming, and matter will accrete through collimated streams infalling onto the magnetic poles of the WD (see e.g. \citealp{Zhilkin_2012} for details of this process). \

Of the CVs that harbour accretion disks, many undergo recurring outbursts known as dwarf novae (DNe; \citealp{Meyer_1981, Osaki_1989}). These events are triggered by thermal instabilities in the accretion disk. As matter accumulates in the disk and heats, the temperature in some locations can cross a critical value, triggering an instability. This produces a heating front that propagates through the disk ultimately resulting in the observed outburst, during which the mass transfer rate onto the WD temporarily increases. This process is explained in detail by the disk instability model (DIM; see \citealp{Lasota_2001} for an in depth review; see also \citealp{Dubus_2018} and \citealp{Hameury_2019} for more recent discussions). Since many CVs are known at relatively small distances, CVs and DNe are excellent environments in which to study and test accretion physics, which is important for several other areas in Galactic (e.g. X-ray binaries) and extragalactic (e.g. Active Galactic Nuclei) astrophysics. \

Various methods haven been employed to search for and characterise CVs. CV candidates can be selected by their blue colour (because accreting sources emit a higher fraction of their light in the blue compared to non-accreting stellar sources; see e.g. \citealp{Wevers_2017} and \citealp{Szkody_2020}), through H$\alpha$ excess (see e.g. \citealp{Rivera_2018}), or by their X-ray emission (see e.g. \citealp{Halpern_2021}). Robust classifications are usually then obtained with spectroscopy, typically via the detection of characteristic emission lines, both in the hydrogen Balmer series and in helium (though note that during outburst, emission lines in CV spectra may weaken and even disappear, and some absorption features may emerge; see e.g. \citealp{Sheets_2007}, \citealp{Breedt_2014}), as well as an overall flat or blue spectral shape. \ 

A key feature of many CVs that allows not only their identification but also the study of their properties in detail is their DN outbursts. During a DN, the brightness of a CV increases by 2--6 optical magnitudes (although very high amplitude outbursts exhibiting brightenings of up to 9 magnitudes have been observed, see \citealp{Tampo_2020} and \citealp{Kawash_2021}). The outbursts typically last days to weeks, and have recurrence times that can vary from days to several years \citep{Lasota_2001, Coleman_2016} or even decades (see e.g. \citealp{Breedt_2014}). Studying DN outbursts is necessary in order to obtain a comprehensive understanding of the accretion processes occurring, so variability studies that can detect DNe are key to the study of CVs.\

The very rapid advances in time-domain astronomy in the last few decades have led to many new detections of DNe, in particular by optical transient surveys such as the Zwicky Transient Facility (ZTF; \citealp{Bellm_2019}), the All Sky Automated Survey for SuperNovae (ASAS-SN; \citealp{Shappee_2014}), the Mobile Astronomical System of TElescope Robots (MASTER; \citealp{Lipunov_2010}), and the Catalina Real-Time Transient Survey (CRTS; \citealp{Djorgovski_2011}). For examples of studies including CV outbursts with these facilities, we refer the reader to \citet{Szkody_2020} for ZTF, \citet{Jayasinghe_2019} for ASAS-SN, and \citet{Drake_2014} for CRTS. Overall, the increasing number of DNe discovered has resulted in a large sample of known systems, which is useful for understanding the overall population and general statistical properties of CVs and their outbursts. \ 
However, some aspects of the accretion physics underlying these systems remains to be explained, and we are still lacking a full, detailed picture of their behaviour during both outburst and quiescence. For example, the standard disk model, that can in principle be used to describe disk-dominated systems such as non-magnetic CVs, has been found to be inconsistent with observations of many CVs (see e.g. \citealp{Godon_2017}). In one ultraviolet (UV) study of CV accretion disks, the overall shapes of theoretical UV spectra were found to be significantly bluer than for many observed systems \citep{Puebla_2007}. They suggest modifications to the disk model that include revising the temperature profile in the inner parts of the disk and the viscosity parameter, as well as considering including an optically thin disk wind. Some studies (see e.g. \citealp{Linnell_2005}) have found better agreement in the UV continuum when modifying the disk model to truncate the inner hot disk. Since this region emits strongly in the UV, truncating the inner disk causes the UV continuum of the model to become shallower, as seen in observations. Characterising the UV emission of DNe, both during outburst and quiescence, is therefore important in understanding to what extent the standard disk model correctly describes CVs. UV observations can lead to modifications of the model that more accurately describe parameters of the systems. \ 

There are also complications remaining in the question of the delay between the onset of the UV and the optical emission during DNe (the UV may rise hours to days after the optical; see \citealp{King_1997,Smak_1998,Stehle_1999,Lasota_2001}). Outbursts triggered by instabilities near the inner accretion disk will result in an `inside-out' outburst, where the heating front propagates outwards through the accretion disk. When the thermal instability occurs in an outer region of the disk, an `outside-in' outburst is produced. The optical emission arises from the overall disk, so the optical rise is observed as soon as the heating wave begins propagating. On the other hand, the UV emission emanates from the inner regions of the accretion disk and from the boundary layer (i.e. it is sensitive to temperatures above 40,000 K; \citealp{Stehle_1999}), so the rise in the UV is observed only when the accretion rate on the WD increases significantly, which happens when the heating front reaches the inner disk. Thus, one might expect a long UV delay in outside-in outbursts, and a short UV delay for inside-out outbursts. However, it is now understood that inside-out outburst should also exhibit significant UV delays: during quiescence, the inner accretion disk (up to a few WD radii) is diluted \citep{Stehle_1999,Schreiber_2003}. Before accretion onto the WD can increase significantly (and thus produce the UV emission), this region needs to be refilled, which occurs on the viscous diffusion time scale, causing a UV delay with respect to the optical. Contrary to initial predictions, the UV delay may actually be greater for inside-out than for outside-in outbursts, because the outside-in heating front propagates faster than an inside-out heating front (see e.g. \citealp{Schreiber_2003}).\ 

Although models based on this description are consistent with many observations, the extent of the UV delay for both types of outbursts is still not fully understood, and for some observed outbursts the UV emission is not in good agreement with predictions from models (see e.g. \citealp{Schreiber_2003}). Importantly, the type of outburst is sensitive to physical parameters of the systems (e.g. mass transfer rate, viscosity), so characterising the UV delays and determining the type of outburst is necessary in the context of studying accretion in CVs. However, only a few systems exist for which simultaneous UV and optical observations exist during the onset of the outburst. A larger sample of both types of DN outbursts with simultaneous UV and optical coverage is necessary to answer the remaining questions.\

The boundary layer (BL; the transition region between the WD and the accretion disk) has proven particularly difficult to model accurately. While the UV flux from CV systems dominates emission from the inner accretion disk and the WD, X-ray flux dominates emission from the BL, with some (extreme-)UV contribution. At low accretion rates (i.e. during quiescence), the BL is expected to be optically thin and emit hard X-rays and produce little EUV emission. At high accretion rates (e.g. during DN outbursts), the BL is optically thick and expected to produce more soft X-ray and EUV photons, as it cools more efficiently so it does not reach the high temperatures needed to produce hard X-rays (\citealp{Popham_1999}, \citealp{Godon_2017} and references therein). However, many X-ray observations do not fully agree with these predictions, as several high-accretion state systems have been observed producing optically thin, soft X-rays (see e.g. \citealp{Baskill_2005,Balman_2014}). These observations may suggest that the BL is indeed optically thin even during outburst, which could be explained by a truncated accretion disk; this would be consistent with the shallow slope in the UV continuum observed in several CVs (discussed above). \

It is therefore apparent that while a general overall picture of the mechanisms underlying CVs and DN outbursts is established, details that are important for the study of accretion physics remain unresolved. UV observations, particularly in combination with (quasi-)simultaneous optical and X-ray data, will help to shed light on these outstanding questions. \ 

\begin{figure*}[t]
    \centering
    \includegraphics[width=0.9\textwidth]{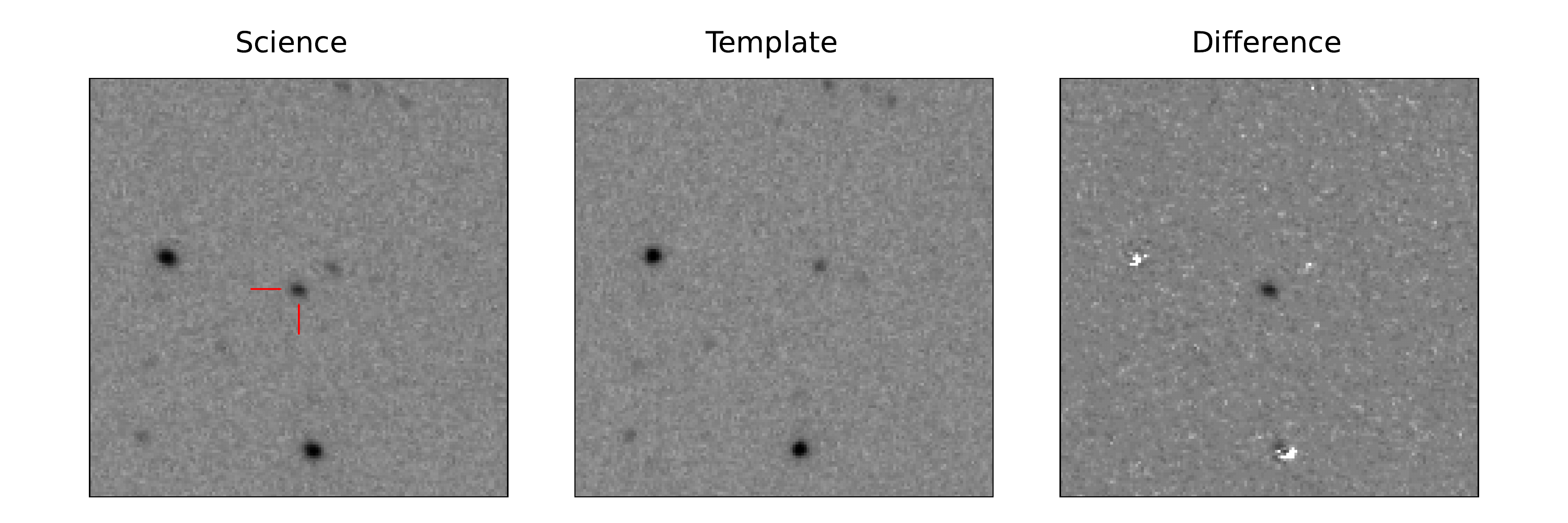}
    \caption{Discovery science image of TUVO-21acq (left), the template image used in the image subtraction process (middle), and the resulting difference (right). Images are 150" by 150". North is up and east is left. The red lines in the science image indicate the position of the transient. The Observation ID (ObsID) of the science image is 00013908006 (extension 2); the ObsID of the template image is 00084458002 (extension 1). The artefacts visible in the difference image result from the image subtraction process (see \citealp{Modiano_2022} for details). Photometry is only performed on the science and template images; difference images are only used for detection of transients.}
    \label{fig:tuvo_discovery}
\end{figure*}

Recently, we have initiated the Transient UV Objects Project (TUVO; see Wijnands et al., in prep), with the main aim of discovering and studying transient and highly variable sources in the UV. This undertaking of TUVO is achieved through \texttt{TUVOpipe} (\citealp{Modiano_2022}), a purposely-built pipeline we developed that uses the image subtraction method to search for transients in the UV using the data obtained with the Ultraviolet Optical Telescope (UVOT\footnote{See \url{https://swift.gsfc.nasa.gov/about_swift/uvot_desc.html} for a description of the instrument, including information about the UV and optical filters available}; see \citealp{Roming_2005,Breeveld_2010}) aboard the \textit{Neil Gehrels Swift Observatory} (\textit{Swift}; \citealp{Gehrels_2004}). With this method we find serendipitous UV transients in near real-time every day (i.e. due to all \textit{Swift} data being public and accessible, with this method we are notified of detected transients within a few hours to a day after the observations are performed), and we follow-up the most interesting candidates with ground- and space-based spectroscopic and photometric observations (see \citealp{Modiano_2022} for how we determine whether a transient candidate is interesting). Due to the prevalence of DNe in the Galaxy and their large outburst amplitudes in the UV, the TUVO project is ideally suited to the discovery of such sources and the study of their outbursts. Many of the UV transients we find indeed can be identified as previously known DNe (see e.g. \citealp{Verberne_2020} and \citealp{Modiano_2022} for some examples). However, some transients we detect may be DNe but lack any, or clear, previous classifications (see e.g. \citealp{ATEL_CV_modiano}).\

In this paper we discuss a UV transient that we discovered as part of the TUVO project in February 2021, and that we have denoted TUVO-21acq. In Sect. \ref{observations} we discuss the observations obtained for this source, including the discovery data, additional archival UV and X-ray observations, and spectroscopic follow-up observations. We also present the information we extracted from the data, from which we were able to securely classify the source as a DN CV, and some of the key properties of the outburst. In Sect. \ref{discussion} we discuss the importance of UV observations in the study of accretion in CVs, in particular in the context of the UV delay and physical characteristics of the accretion disk. We also discuss this work as a proof-of-concept for the TUVO project.

\section{Observations, analysis, and results}  \label{observations}

\subsection{Discovery}  \label{discovery}
\begin{figure*}[h]
    \centering
    \includegraphics[width=0.85\textwidth]{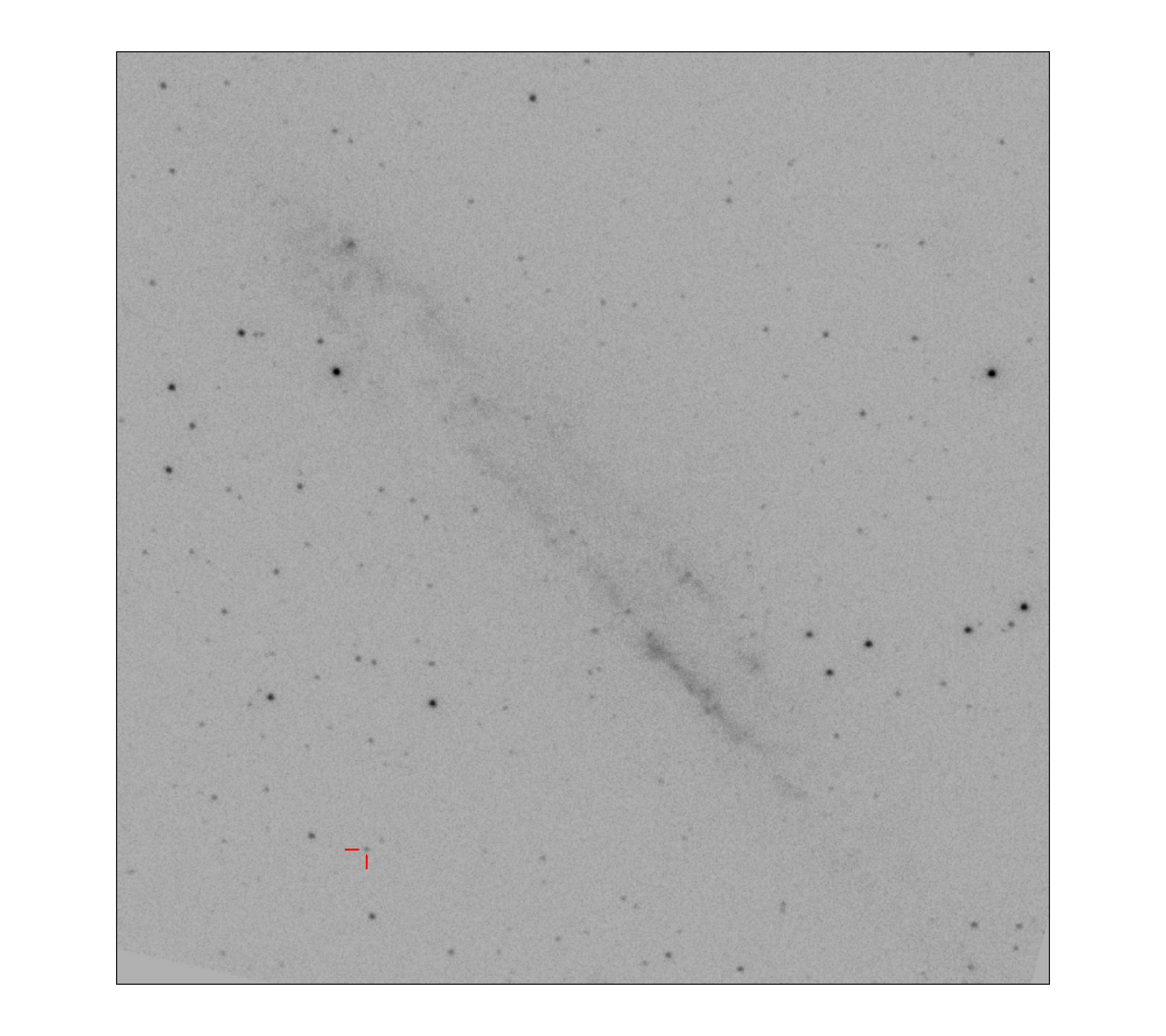}
    \caption{One of the UVW1 images in which TUVO-21acq was detected. The spiral galaxy NGC4945 is clearly visible. The red lines indicate the position of TUVO-21acq. The image is 13' by 13'. North is up and east is left. The ObsID of the image is 00013908006 (extension 1).}
    \label{fig:full_image}
\end{figure*}

TUVO-21acq was first detected in UVOT data on 23 February 2021, during our daily running of \texttt{TUVOpipe} (see \citealp{Modiano_2022} for details of the pipeline and how it detects transients). The transient was found in the field that targeted the spiral galaxy NGC 4945, and was located 6' from the centre of that galaxy. In Fig. \ref{fig:tuvo_discovery} we show stamps of the discovery image (left panel) alongside the template image (middle panel), and the difference image (produced by the TUVO pipeline via image subtraction with \texttt{hotpants}\footnote{\tiny\url{https://github.com/acbecker/hotpants}}; \citealp{Becker_2015}; right panel). The full discovery image is shown in Fig. \ref{fig:full_image}. Astrometric errors in UVOT images may be up to a few arcseconds (see \citealp{Poole_2008,Breeveld_2010}), so to obtain the best possible position for the source we first used \url{astrometry.net} to determine the astrometric solution of the discovery image to a precision of 0.5". We then searched the Gaia \citep{Gaia_2016} DR3 catalogue \citep{Gaia_dr3} for a source at this position, and found a match at Right Ascension=13:05:44.72, Declination=-49:32:58.3 (J2000; source ID 6084567987998395520), which was 0.12" from the UVOT position we determined. We therefore report the Gaia coordinates as the best possible position of TUVO-21acq.\

\subsection{Further UVOT observations}  \label{uvot_archival_data}

\begin{figure*}[h]
    \centering
    \includegraphics[width=0.9\textwidth]{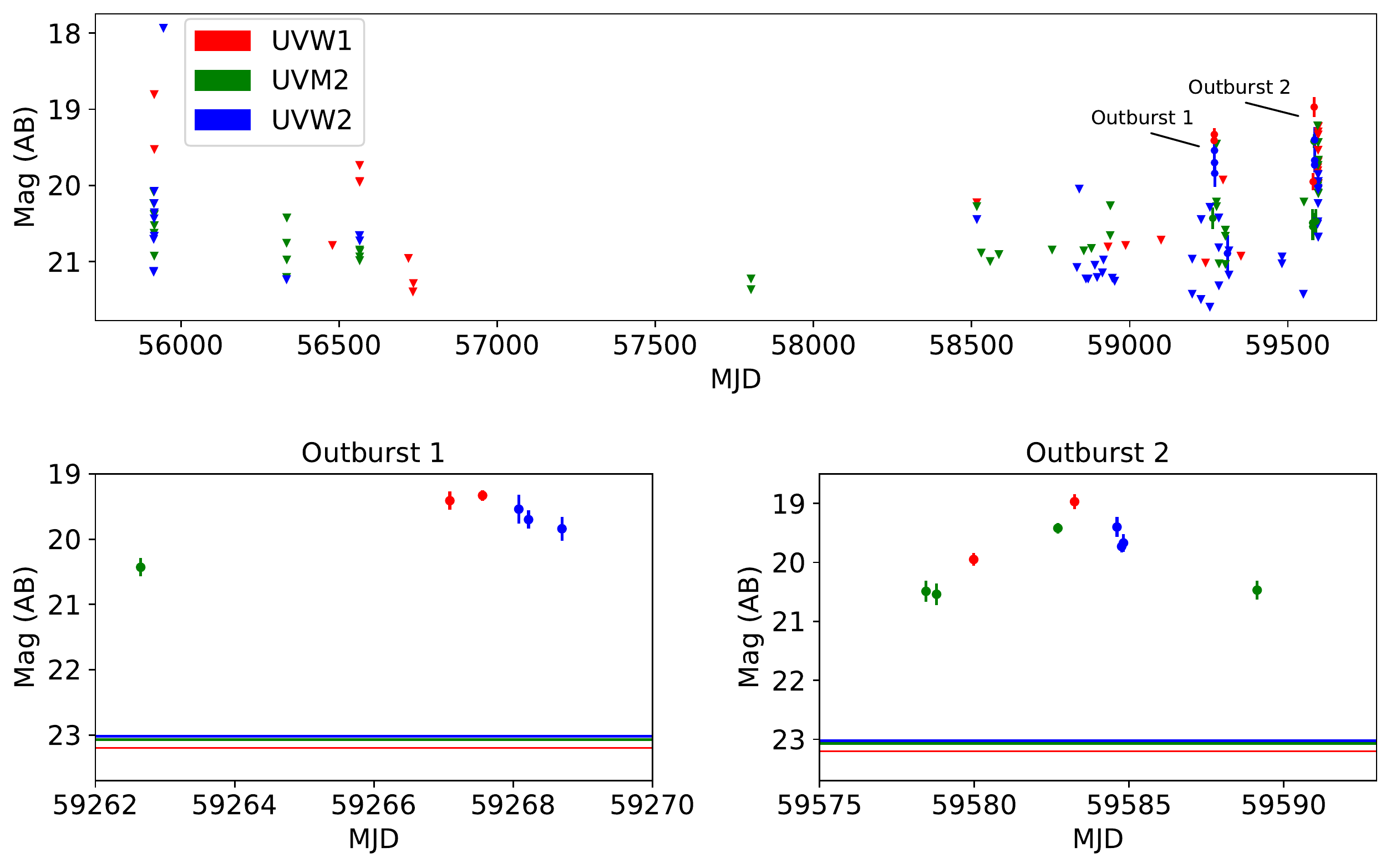}
    \caption{UVOT light curves of TUVO-21acq. Upper panel: long-term light curve, created using all UVOT data covering the position of the source in all three UV filters. Data from optical filters are not shown for aesthetic purposes, since very few images were obtained in these filters and only weak upper limits were found (i.e. no detections). Lower panels: zoomed-in views of the two outburst segments that are labelled in the top panel. Horizontal lines indicate the quiescence level (see Sect. \ref{outburst_amplitude}). Triangles indicate upper limits (indicating that the source was not detected at $\ge3\sigma$).}
    \label{fig:pv_lightcurve}
\end{figure*}

Almost a year after the first outburst, \texttt{TUVOpipe} made a second detection of this transient from an observation carried out on 04 January 2022 in the UVW1 filter (although the outburst had started at least as early as 31 December 2021 as detected in a UVM2 image), indicating that TUVO-21acq underwent at least two outbursts. In order to check for any additional past outburst or variable behaviour, we searched for any previous detections in all archival UVOT data. We downloaded all available images with pointings less than 15'\footnote{Due to the field of view of UVOT (17'x17') and the variable roll angle used for observations, a 15' search radius always ensures that we use all UVOT data covering the source position} from the position of our source, using all data from the full UVOT archive, which is hosted at the High Energy Astrophysics Science Archive Research Centre (HEASARC)\footnote{\tiny\url{https://heasarc.gsfc.nasa.gov/cgi-bin/W3Browse/swift.pl}}. This resulted in a total of 160 images, with the first UVOT observation covering this position taken on 18 December 2011, and of irregular sampling cadence (see Figure \ref{fig:pv_lightcurve}). The field has been observed primarily using the three UV filters available for UVOT, with 9, 9, 9, 30, 48, and 55 images obtained in the filters V, B, U, UVW1, UVM2, UVW2, respectively. \ 

We aligned all UVOT images to the discovery image of the first outburst using the \texttt{register\_translation} function of the \texttt{scikit-image}\footnote{\tiny\url{https://scikit-image.org/}} package. We then performed photometry at the source location on every image, using the standard UVOT photometry tool \texttt{uvotsource}\footnote{\tiny\url{https://heasarc.gsfc.nasa.gov/lheasoft/ftools/headas/uvotsource.html}}. To compute the source flux, \texttt{uvotsource} requires as input a source and background region. As source region we used a circle centred on the source position and with radius 7.0", to ensure that the wings of the brightness profile of the source were always included in the source region. The background region was defined as four circles of radius 15", and were selected so as to be nearby the source, on different sides of the source (to minimise background gradient effects, e.g. due to the nearby galaxy) and devoid of contaminating sources.\ 

The resulting long-term light curve is shown in Fig. \ref{fig:pv_lightcurve} (top panel). The source was only detected in a total of 15 images, all obtained using the three UV filters. From the light curve it is clear that these detections are grouped into two independent outbursts. Zoomed-in views of both outbursts are shown in the lower panels of Fig. \ref{fig:pv_lightcurve}. To distinguish between detections and non-detections (i.e. upper limits in the light curve), \texttt{uvotsource} requires a user-selected significance as input, for which we used 3-$\sigma$. To verify the absence of any weak detections in the data, we also ran \texttt{uvotsource} on all images with 1-$\sigma$. No detections were made in any images in which \texttt{uvotsource} did not make a 3-$\sigma$ detection.\

\subsection{Photometric analysis} \label{lightcurve_analysis}

\subsubsection{Outburst amplitudes} \label{outburst_amplitude}

The light curve shows that the observations occurred during the rise, peak, and decay stages of both outbursts. In both outbursts, the UVW1 observations likely represent the approximate peak, because since accreting sources are expected to display flat spectra, the brightness at any stage of the outburst should be very similar between the three UV bands. We note here the potential caveat that due to the interstellar absorption bump at 2175 $\AA$ \citep{Fitzpatrick_1988}, the UVM2 flux is more absorbed than the UVW2 and UVW1 fluxes, and can thus be underestimated. However, given the quiescent broad-band UV spectrum observed, which is consistent with a flat shape (see Table \ref{tab:photometry}), we expect this effect to be minimal, since no dip is observed in the UVM2 band.\

\begin{figure}[h]
    \centering
    \includegraphics[width=0.35\textwidth]{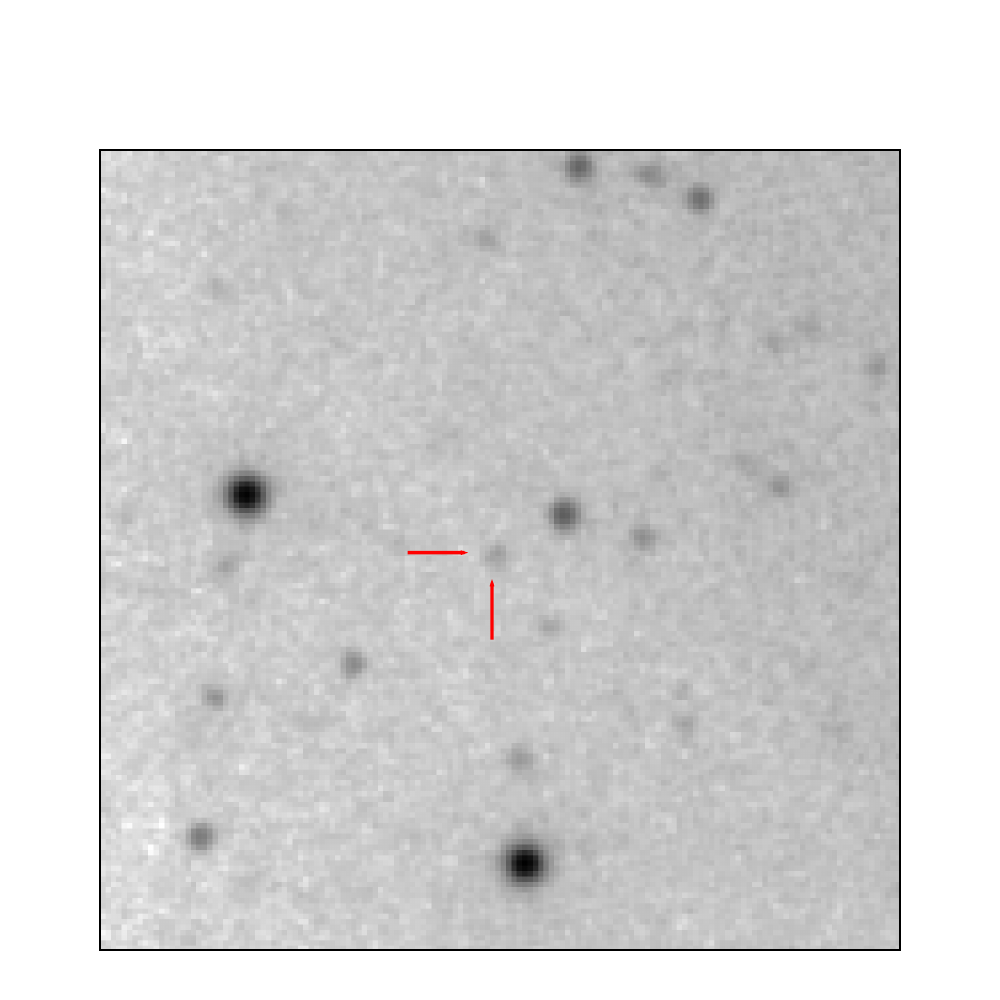}
    \caption{Summed UVW1 image showing the detection of TUVO-21acq during quiescence. The image was created by co-adding all archival UVW1 exposures covering the position of TUVO-21acq while the source was not in outburst (i.e. images taken during the outbursts were omitted). The total exposure time is 42 ks. The image is 150" by 150". North is up and east is left. The red lines indicate the position of TUVO-21acq.}
    \label{fig:summed_image}
\end{figure}

To better constrain the amplitude of the outbursts, knowledge of the source brightness during quiescence, or at least a more stringent upper limit, is necessary. To obtain this, we co-added the archival UVOT images from all the observations obtained during quiescence (i.e. we excluded the data taken during the two outbursts) for each of the three UV filters independently. This was done using the HEASARC tools \texttt{fappend}\footnote{\tiny\url{https://heasarc.gsfc.nasa.gov/lheasoft/ftools/fhelp/fappend.txt}} and \texttt{uvotimsum}\footnote{\tiny\url{https://heasarc.gsfc.nasa.gov/lheasoft/ftools/headas/uvotimsum.html}}. The summed UVW1 image is shown in Fig. \ref{fig:summed_image}. The source was detected in each of the summed images for the three filters, so we ran \texttt{uvotsource} on the summed images, using the same source and background regions used previously. The resulting magnitudes in the three UV filters indicated outburst amplitudes between 2.4 and 4.4 mags (see Table \ref{tab:photometry}). We note, however, that this method of determining the outburst amplitude assumes the brightness of the source to be constant during quiescence. Flux from CVs may in fact vary during quiescence. For example, quiescent CV variability known as flickering \citep{Bruch_1992} exhibits amplitudes ranging from millimagnitudes up to $\sim$1 mag in the optical, with the bluer and UV bands typically displaying larger amplitudes than the redder optical bands (see e.g. \citealp{Bruch_2015,Bruch_2021}). However, in this case the lack of detections in individual images during quiescence and the sampling cadence mean that it is not possible to estimate the exact quiescent levels immediately prior and subsequent to the outbursts. Therefore, the exact amplitudes quoted must be taken with a grain of salt, but they are nonetheless highly indicative of the true amplitude of the brightening caused by the outbursts. \ 

\begin{table*}[t]
    \centering
    \begin{tabular}{P{2cm} P{2cm} | P{2cm} P{2cm}| P{2cm} P{2cm}}
        \multicolumn{1}{c}{} & \multicolumn{1}{c}{} & \multicolumn{2}{c}{Outburst 1} & \multicolumn{2}{c}{Outburst 2} \\
       \hline
        Filter & Quiescence & Peak & Amplitude & Peak & Amplitude \\
        \hline \hline
        UVW1 & $23.4\pm0.3$ & $19.3\pm0.1$ & >4.1 & $19.0\pm0.1$& >4.4 \\
        UVM2 & $22.8\pm0.2$ & $20.4\pm0.1$ & >2.4 & $19.4\pm0.1$ & >3.4 \\
        UVW2 & $23.0\pm0.1$ & $19.5\pm0.2$ & >3.5 & $19.4\pm0.2$ & >3.6 \\ \hline

    \end{tabular}
    \caption{Photometric magnitudes measured for TUVO-21acq in the three UV bands used during the UVOT observations. All magnitudes are given in the AB system. Quiescence measurements were obtained using the summed images, which exclude images obtained during the outbursts. The errors are the 1$\sigma$ statistical errors on the photometry added in quadrature with the systematic UVOT calibration errors, as given by \texttt{uvotsource}. The outburst amplitudes are lower limits because only the UWV1 detections are likely to have occurred around the outburst peak (see Fig. \ref{fig:pv_lightcurve}).}
    \label{tab:photometry}
\end{table*}

In Table~\ref{tab:photometry} we provide a summary of the photometric measurements made during both outbursts and in quiescence, and the derived constraints on the amplitude of the outbursts. The UVM2 brightening of the first outburst is significantly weaker than all other derived amplitudes, likely because the only detection during outburst in this filter was made several days before the peak (see Figure \ref{fig:pv_lightcurve}). Typical DNe brighten by 2--6 magnitudes in the optical (\citealp{Coleman_2016,Dobrotka_2006,Breedt_2014, Belloni_2019}), so the outburst amplitudes we measure for both observed outbursts of TUVO-21acq are consistent with a DN (note that the UV amplitude may be higher in the UV for some systems; see e.g. \citealp{Giovannelli_2008,Neustroev_2017}). \

\subsubsection{Outburst timescales} \label{outburst_timescales}

We obtained some constraints on the outburst durations of TUVO-21acq using all the UVOT data covering both outbursts. By calculating the time between the first and last detections of each outburst, we determined lower limits for the first and second outburst durations of 6.1 and 10.7 days, respectively. We also obtained upper limits for the duration by calculating the time between the latest non-detection before the outburst and the first non-detection after the outburst, on the assumption that during any non-detection the system was not in outburst. The upper limits we estimated for the outburst durations are 21.1 and 45.0 days, respectively for the first and second outbursts. These overall duration ranges for TUVO-21acq outbursts (6.1--45.0 days) are consistent with typical DN durations (2-20 days; \citealp{Lasota_2001}, \citealp{Breedt_2014}; note that so-called superoutbursts; \citealp{Osaki_1989_sup}, may last up to several weeks). The light curves of the two outbursts (see Fig. \ref{fig:pv_lightcurve}, lower panels) exhibit fairly similar behaviour in terms of the shape and timescales of both the rise and decay stages. \

Due to the non-uniform cadence of the UVOT observations and the fact that there are only two known outbursts of this source, it is difficult to make a precise estimate of the recurrence time. However, an upper limit for the recurrence time can be found simply by measuring the time between the two known outbursts. Using the outburst peaks mentioned, we calculated an upper limit for the recurrence time of 316 days. This is a weak constraint, however, because as shown clearly by the gaps in the light curve and the duration of the outbursts, several outbursts could have been missed. The upper limit is nonetheless consistent with DN recurrence times, which range from a few days to several years to even decades \citep{Lasota_2001, Breedt_2014, Belloni_2019}. These recurrence times can vary significantly between different sub-classes of DNe \citep{Breedt_2014}; and they may also vary considerably even for individual sources (i.e. a single system may not necessarily have a constant recurrence time). \ 

We also estimated the duty cycle of TUVO-21acq (the fraction of time it spends in outburst). The duty cycle is given by ratio of the outburst duration to the recurrence time. Using the maximum and minimum durations we measure for both outbursts and the upper limit obtained for the recurrence time, we found a duty cycle of 0.02--0.14. Since we used an upper limit for the recurrence time, this represents a lower limit for the duty cycle. A simplistic estimate of the duty cycle can also be obtained by the ratio of the number of observations in which the source was detected (i.e., when it was in outburst) over the number of observations in which it was not detected (i.e., when it was in quiescence). This calculation yielded 0.09. However, it comes with the assumption that the sampling cadence is random and that all observations are independent of each other. This is not the case for the UVOT observations, so for example, several sets of consecutive observations may be probing the same quiescent phase. This results in skewing the value of the duty cycle obtained. The approximate values we obtained for the duty cycle of TUVO-21acq using both methods are nonetheless similar, and consistent with the range of DN duty cycles determined by \citet{Coppejans_2016} from several hundred DNe observed in the CRTS of 0.01-0.36.\ 

\subsection{XRT} \label{xrt_detection}

One of the primary features of \textit{Swift} is the simultaneous UV or optical and X-ray coverage, and most observations are performed with both the UVOT and the X-ray telescope (XRT; with an energy range of 0.3-10 keV; \citealp{XRT_2005}) pointing at the same field. We checked the XRT data of TUVO-21acq for any X-ray emission from the source both during the outburst and in quiescence. There are 153.6 ks of XRT data available covering the position of the source in photon-counting (PC) mode. 0.6 ks of data are also available in windowed-timing mode but due to the low count rates for our source (see below), we only report on the PC data.\ 

Using the online XRT analysis tool\footnote{\tiny\url{https://www.swift.ac.uk/user_objects/}}\citep{Evans_2009}, we extracted combined XRT images of all the exposures taken during outburst, that is, including only the ObsIDs in which there was a UVOT detection of the source. We did this independently for both outbursts and also combining the two outbursts (see reproduction package for details of the ObsIDs used for each selection). There were 4.4 ks and 5.1 ks of exposure time available for the first and second outbursts, respectively. For the individual outbursts and for the combination of both, there was no XRT detection on the source position (using the Gaia position). To obtain an upper limit, we used the HEASARC tool Webpimms\footnote{\tiny\url{https://heasarc.gsfc.nasa.gov/cgi-bin/Tools/w3pimms/w3pimms.pl}}, assuming an N${_H}$ value of 1.4$\times$10$^{22}$ (using the predicted N${_H}$ on the source position using the HEASARC tool\footnote{\tiny\url{https://heasarc.gsfc.nasa.gov/cgi-bin/Tools/w3nh/w3nh.pl}}), a spectrum with a power-law index of 1.8 (a typical value for non-magnetic quiescent CVs; see e.g. \citealp{Maiolino_2020}), and the 3-$\sigma$ confidence limit prescriptions from \citet{Gehrels_1986}. We refer to the reproduction package of this paper for the full calculation. We report XRT uppers limit on the unabsorbed fluxes in the 0.3-10 keV band of 1.9$\times$10$^{-13}$ erg cm$^{-2}$ s$^{-1}$ and 1.7$\times$10$^{-13}$ erg cm$^{-2}$ s$^{-1}$ for the first and second outbursts, respectively. \ 

Using the online XRT analysis tool, we then extracted the combined XRT image of all the exposures taken during quiescence, therefore omitting all the ObsIDs in which there was a UVOT detection of the source. This resulted in a total of 144.1 ks of exposure. In this combined quiescent image, using the HEASoft\footnote{\tiny\url{https://heasarc.gsfc.nasa.gov/docs/software/lheasoft/}} \texttt{ximage}\footnote{\tiny\url{https://www.swift.ac.uk/analysis/xrt/xrtcentroid.php}} we found a detection at the position of TUVO-21acq with a count rate (in the energy range 0.3--10 keV) of 4.9$\times$10$^{-4}$ $\pm$ 8.5$\times$10$^{-5}$ counts per second (with a signal-to-noise ratio 5.7).\ 

\begin{figure}[h]
    \centering
    \includegraphics[width=0.45\textwidth]{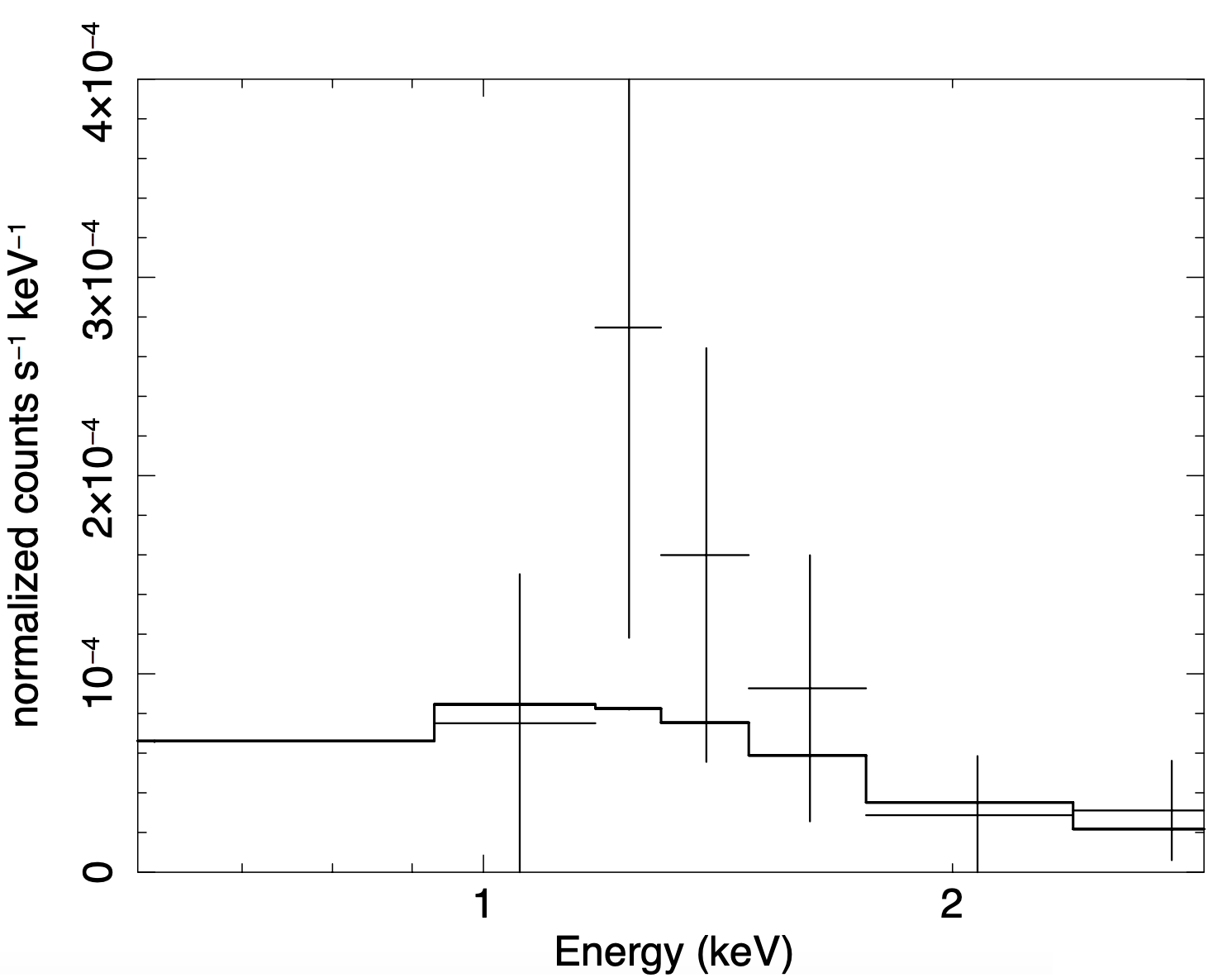}
    \caption{Fitted X-ray spectrum extracted from combined quiescence XRT image of TUVO-21acq. The bold black line is the best fit using the Mekal model.}
    \label{fig:xrt_spec}
\end{figure}

Although the source count rate is low, there were enough photons to extract a spectrum from this quiescence image (see Fig. \ref{fig:xrt_spec}); this was also done using the online XRT tool (which produces the source and background spectra as well as the ancillary files). Using the HEASoft tool \texttt{grppha}\footnote{\tiny\url{https://heasarc.gsfc.nasa.gov/lheasoft/ftools/fhelp/grppha.txt}} we rebinned the extracted spectrum with 5 photons per bin. Using XSPEC \citep{Arnaud_1996}, we fitted the spectrum in order to extract the flux and obtain an indication of the shape of the quiescent spectrum. Due to the low number of photons, it was only possible to use simple models, and we used the C-statistic (instead of $\chi^{2}$), which is appropriate for faint sources. We used the Mekal model (\citealp{Mewe_1985}, that has been used to fit quiescent CV X-ray spectra; see e.g. \citealp{Heinke_2006}), using a fixed column density N${_H}$ of 1.4$\times$10$^{22}$ (as also used above), a fixed redshift of 0, and leaving other parameters free (plasma temperature, metal abundance, and a normalisation parameter). From the fitted spectrum we extracted a flux of 1.11$\times$10$^{-14}$ erg cm$^{-2}$ s$^{-1}$ in the range 0.3-10 keV. The fitted spectrum is shown in Fig. \ref{fig:xrt_spec}.\ 

\subsection{External archival information} \label{archival_info}

To obtain information available about any known counterparts of this source, we cross-matched its position with astronomical databases, both variable star and transient catalogues as well as persistent source catalogues (i.e. deep sky surveys, that may have observed the source while it was in quiescence). We refer to \citet{Modiano_2022} for a full list of all the catalogues we probe in the TUVO project and further information for each. None of the transient or variable catalogues revealed a previously known source within 5" of the position of TUVO-21acq, suggesting that this source was not previously known to be variable. However, a useful match was found in the Gaia catalogue (see Sect. \ref{discovery}). The Gaia photometric measurements in the broad $G$, blue $G_{BP}$ and red $G_{RP}$ bands\footnote{See \url{https://www.cosmos.esa.int/web/gaia/iow_20180316}.} are 20.8, 20.4, and 20.1 magnitudes, respectively. The $G_{BP}-G_{RP}$ colour of $\sim$0.3 is consistent with the relatively blue colour expected from CVs (see e.g. \citealp{Abril_2020}). No reliable Gaia distance is available for the source, so we are not able to determine absolute magnitudes or luminosities for TUVO-21acq.

\subsection{SALT spectrum} \label{salt}

\begin{figure*}[t]
    \centering
    \includegraphics[scale=0.6]{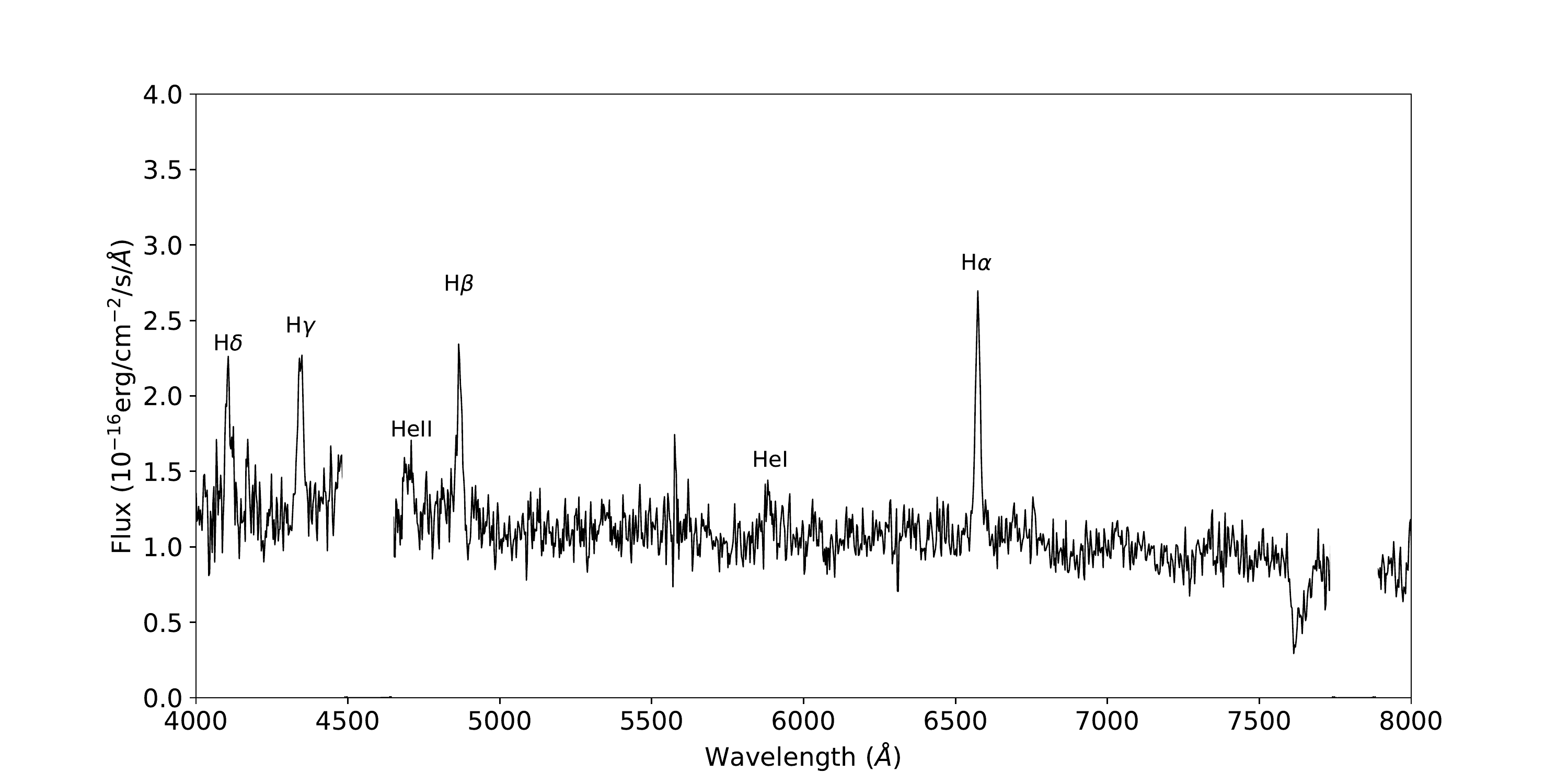}
    \caption{SALT-RSS spectrum of TUVO-21acq. The spectrum is wavelength calibrated, background subtracted and with relative flux calibrations applied. Emission lines from hydrogen Balmer series and from helium are labelled. The gaps at $\sim$4500 $\AA$ and $\sim$7800 $\AA$ represent gaps in the chip. The emission feature at $\sim$5577 $\AA$ and the absorption feature at $\sim$7600 $\AA$ are both atmospheric features that are not fully removed by the flux calibration. The entire spectrum is shifted towards the red by 2.0 $\AA$ as this was the error we found on the wavelength calibration by using the 5577.3 atmospheric oxygen line.}
    \label{fig:spectrum}
\end{figure*}

We additionally carried out follow-up spectroscopic optical observations with the Southern African Large Telescope (SALT; \citealp{Buckley2006}) at Sutherland Observatory. A 1200s exposure was obtained with the Robert Stobie Spectrograph (RSS; \citealp{Burgh2003}) on 09 April 2021, starting at 01:29 UTC. We used the PG300 grating, which covers the range 3700-8300 $\AA$ (with gaps in the chip at 4480-4645 $\AA$ and 7740-7880 $\AA$) and a slit width of 1.25", resulting in a mean resolving power of R$\sim$420 (14.8 $\AA$ resolution). Sky conditions were clear, with seeing $\sim$1.0". \

To reduce the spectra we used the PyRAF-based PySALT package \citep{Crawford2010}\footnote{\tiny\url{https://astronomers.salt.ac.za/software/pysalt-documentation/}}, which includes corrections for gain and cross-talk, and performs bias subtraction. We extracted the science spectrum using standard
IRAF\footnote{\tiny\url{https://iraf.net/}} tasks, including wavelength calibration (two Argon calibration lamp exposures were taken, one immediately before and one immediately after the science spectrum), background subtraction, and 1D spectra extraction. Due to the SALT design, absolute flux calibration is not possible\footnote{The pupil (i.e. the view of the mirror from the tracker) moves during all SALT observations, causing the effective area of the telescope to change during exposures. Therefore, absolute flux calibration cannot be done. See \citet{Buckley2006} and \citet{Crawford2010} for details.}. However, by observing spectrophotometric standards during twilight \citep{Buckley2018}, we were able to obtain relative flux calibration, allowing recovery of the correct spectral shape and relative line strengths. The reduced spectrum is displayed in Figure \ref{fig:spectrum}.\

We then analysed the reduced spectrum using tools within the \texttt{specutils}\footnote{\tiny\url{https://specutils.readthedocs.io/en/stable/}} package available in \texttt{astropy}\footnote{\tiny\url{https://www.astropy.org/}} for analysing spectroscopic data. To ensure the best possible wavelength calibration (that intrinsically had an error of $\sim$3 $\AA$ from the instrument and reduction), we first determined the centroid wavelength of the 5577.3 $\AA$ atmospheric oxygen line in one of the calibration spectra. We found it to be at 5575.3 $\AA$, and therefore shifted the entire reduced spectrum by 2.0 $\AA$ (the difference in error on the wavelength calibration for different parts of the spectrum is negligible). We extracted basic parameters of the key spectral features, focusing on the primary emission lines clearly visible (see Fig. \ref{fig:spectrum}): H$\alpha$ $\lambda$6563, H$\beta$ $\lambda$4861, H$\gamma$ $\lambda$4340, H$\delta$ $\lambda$4102, He {\sc ii} $\lambda$4686, and He {\sc i} $\lambda$5876. We first determined the continuum using the \texttt{fit\_generic\_continuum} function of \texttt{specutils}. We divided the full spectrum by the continuum in order to obtain a continuum-normalised spectrum, from which we calculated the EW and Full Width at Half Maxima (FWHM) of each emission line. Centroid wavelengths and relative line fluxes were determined from the continuum-subtracted spectrum, which was created by subtracting the continuum from the full spectrum. The results are shown in Table \ref{tab:spec_lines}. \

\begin{table}[h]
    \small
    \centering
    \begin{tabular}{P{1.5cm}P{1.6cm}P{1.1cm}P{1.1cm}P{1.1cm}}
    \toprule
       Line & Observed centroid ($\AA$) & Relative flux & EW ($\AA$) & FWHM ($\AA$) \\ \hline \hline
        H$\alpha$ $\lambda$6563 & 6571 & 2.4 & -36 & 32  \\
        H$\beta$ $\lambda$4861 & 4862 & 2.0 & -28 & 40  \\
        H$\gamma$ $\lambda$4340 & 4342 & 1.7 & -23 & 50  \\
        H$\delta$ $\lambda$4102 & 4105 & 1.0 & -14 & 36  \\
        He {\sc ii} $\lambda$4686 & 4697 & 1.2 & -16 & 58  \\
        He {\sc i} $\lambda$5876 & 5876 & 0.3 & -9 & 58  \\
    \bottomrule
    \end{tabular}
    \caption{Centroid wavelengths, relative fluxes, equivalent widths (EW), and FWHM of the strongest emission lines visible in the SALT-RSS spectrum of TUVO-21acq. Fluxes are normalised to H$\delta$. Negative EWs indicate emission.}
    \label{tab:spec_lines}
\end{table}

A clear signature of accreting sources in optical spectra is the presence of emission lines when in quiescence, in particular hydrogen Balmer series and helium lines. These lines are clearly observable in the spectrum of TUVO-21acq (Figure \ref{fig:spectrum}). In addition, accretion is usually discernible by a relatively flat or blue overall spectral shape. Therefore, the spectrum is indicative of a system in which the emission is dominated by accretion processes, strengthening the classification of the source as a CV. \ 

The spectrum was obtained when the source was in quiescence. DN outburst spectra typically show significantly weaker, or even entirely absent emission lines with respect to quiescent DNe (see e.g. \citealp{Sheets_2007}, \citealp{Breedt_2014}). Typical spectra of DN systems during quiescence show line flux ratios of He {\sc i} $\lambda$5876/(H$\alpha$) $\sim$0.2-0.4 \citep{Breedt_2014}. We measure the ratio at 0.13, indicating a somewhat diminished helium deficiency with respect to typical quiescent DN CVs.

We measured the radial velocities (RVs) of the emission lines using the centroid wavelength of each line found using \texttt{specutils} (see Table \ref{tab:spec_lines}). We found RVs of all lines to be consistent with each other within the measurement error. Time-resolved, high-resolution spectroscopy of this source is required for a complete, detailed study of the RVs.\

\section{Discussion} \label{discussion}

Within our TUVO project, we have discovered and characterised a new cataclysmic variable in the UV, that we have denoted as TUVO-21acq. The first detection was made in February 2021 using our dedicated pipeline \texttt{TUVOpipe}, which analyses daily UVOT data to look for transients. A second UV outburst was detected in January 2022. Key parameters of the outbursts derived from UVOT photometry (both for the outburst data and on the quiescence data, i.e. using archival images), namely the amplitudes, durations, and recurrence time, are found to be fully consistent with DNe (see Sect. \ref{outburst_amplitude} and \ref{outburst_timescales}). We additionally obtained a spectrum with SALT of the source during quiescence. The overall shape and key emission features exhibited in the spectrum are indicative of a quiescent CV. Therefore, we can provide a robust classification of TUVO-21acq as a DN CV.\ 

In this section we discuss the implications of our findings for this source and in the context of the UV emission of DNe. Despite DNe having been studied for many decades, questions about the detailed accretion processes involved remain. As discussed in Sect. \ref{introduction}, the temperature profiles and exact structure of the accretion disk and boundary layer are not fully understood; moreover, observations of the UV delay (with respect to the optical) have been found to not always agree with theoretical predictions. Here we also examine the importance of searching for transients in the UV wavelength range.

\subsection{UV delay} \label{uv_delay}

As discussed in Sect. \ref{introduction}, questions still remain regarding the UV delay of DNe. In answering line of research, we note that \textit{Swift} is very advantageous. Its rapid pointing capabilities and six filters covering UV and optical ranges mean that (quasi-)simultaneous observations in the UV and optical can be obtained. It is therefore well-suited to studying the UV delay in DNe, since the delays are expected from between a few hours to a few days. Such observational data (i.e. the UV flux and its delay with respect to optical) can then be used as input in disk models in order to obtain more precise estimates of parameters such as the viscosity, mass transfer rate, and truncation radius of the accretion disk. Targeted simultaneous UV and optical observations of known, recurrent DNe during the onset of their outbursts are therefore useful in characterising and understanding accretion processes. While our pipeline itself is unlikely to detect UV delays in DNe serendipitously (the chance of the rise of a DN outburst occurring in the FoV of UVOT during multi-filter observations targeting a different source, though nonzero, is very low), it can discover new, UV-bright DNe that can be optimal targets for targeted studies of UV delays with \textit{Swift} (or other facilities).

\subsection{UV spectral slope} \label{uv_spec_slope}

Ultraviolet spectra of quiescent CVs are important because various parameters of the standard disk model, including the temperature profile in the inner accretion disk and the viscosity parameter, are sensitive to the slope of the UV spectrum \citep{Puebla_2007,Godon_2017}. Measuring the UV spectral slope can therefore help to better constrain some of the key physical characteristics of the accretion disks in CVs. Although with lower precision than when using spectroscopy, in principle this can also be done with broad-band UV spectra, that is, using photometry in several UV bands. For this reason, studies of quiescent CVs in the UV may help to obtain a larger sample of UV spectral slopes of CVs, so as to better study this problem. However, due to the faintness of TUVO-21acq in the quiescence images and the lack of detections in individual images, we cannot obtain a reliable spectral slope based only on the available UVW1, UVM2, and UVW2 photometry for this particular source. However, the relatively flat shape of the UV broad-band spectrum (see Table \ref{tab:photometry}) is consistent with the shapes of typical quiescent CV spectra in the UV (see \citealp{Puebla_2007} for many examples of such spectra).

\subsection{X-rays and the X-ray-optical flux ratio} \label{xray}

The simultaneous XRT data obtained by \textit{Swift} during all observations can help to further study these types of sources by providing some information on physical parameters of the system. Specifically, the ratio of the X-ray to the optical or UV flux of CVs is thought to correlate with orbital periods and mass transfer rates: several studies have found that for quiescent non-magnetic CVs, $F_{X}/F_{opt}$ decreases with increasing mass transfer rate and with increasing orbital period (see \citealp{Patterson_1985,Richman_1996,Kuulkers_2006} and references therein). This relationship is explained primarily by variations in the optical flux, which is thought to increase with increasing mass transfer rate (while the X-ray flux remains fairly constant; see \citealp{Kuulkers_2006}). A relation has been derived empirically between the mass transfer rate $\dot{M}$ and $F_{X}/F_{opt}$ (see \citealp{Richman_1996}). Using the quiescent X-ray flux in the range 0.3-10 keV of 1.11$\times$10$^{-14}$ erg cm$^{-2}$ s$^{-1}$ (see Sect. \ref{xrt_detection}) and the quiescent Gaia $G$-band flux of 6.95$\times$10$^{-14}$ erg cm$^{-2}$ s$^{-1}$ (calculated using the flux given by Gaia in photoelectrons/s and Equation 5.24 in the Gaia documentation\footnote{\url{https://gea.esac.esa.int/archive/documentation/GEDR3/Data_processing/chap_cu5pho/cu5pho_sec_photProc/cu5pho_ssec_photCal.html}}), we obtain a flux ratio $F_{X}/F_{opt}$ of $\sim$0.2. This is consistent with the the higher end of $F_{X}/F_{opt}$ found in quiescent non-magnetic CVs (see e.g. \citealp{vandenBerg_2004,Kuulkers_2006,Balman_2014}), and thus indicates a relatively low mass transfer rate for TUVO-21acq. Using Equation 5 in \citet{Richman_1996}, we obtained log$\dot{M}\sim$ 16 g s$^{-1}$, which is indeed among the lower mass transfer rates for typical non-magnetic CVs described in their paper. \ 

An additional empirical relationship exists between the EW of the H$\beta$ emission line and $F_{X}/F_{opt}$ (to an accuracy of a factor of 3, see \citealp{Kuulkers_2006}). This correlation stems from the fact that the EW (H$\beta$) is a proxy for the optical flux from the disk, which is related to $\dot{M}$. Using the relevant relationship (Equation 4 in \citealp{Richman_1996}) and our measured EW (H$\beta$), we obtain a flux ratio of $\sim$0.3. This is consistent with the value obtained from the previous method ($\sim$0.2) within the accuracy of the empirical relationship used.

\subsection{TUVO proof-of-concept} \label{tuvo_proofofconcept}

An important aim of this study is to serve as a proof-of-concept for the TUVO project (Wijnands et al., in prep) and thus to demonstrate the ability of our dedicated pipeline \texttt{TUVOpipe} \citep{Modiano_2022} in discovering new, previously unclassified transients in the UV wavelength range. The UV transient sky remains to be systematically explored, despite the fact that many types of interesting, highly energetic transient phenomena exhibit strong UV emission. These transients include outbursts from accreting white dwarfs such as TUVO-21acq, but range from flares from active stars through to supernovae and tidal disruption events (see \citealp{Sagiv_2014} for an overview of different types of transients expected to emit strongly in the UV). The UV is a particularly interesting wavelength through which to discover and study these phenomena, because the emission often peaks in the UV, and potentially earlier than at optical wavelengths (see e.g. \citealp{Neustroev_2017} for a DN example; \citealp{Gezari_2015} for a supernova example, and \citealp{Gomez_2020} for a tidal disruption event example). Much of the physics underlying these events is currently not well understood. Searching for transients in the UV regime would allow for potentially crucial early UV data to be obtained, in particular if in combination with (quasi-)simultaneous observations at other wavelengths. Although there are several large-scale UV transient surveying observatories in proposal and development stages (see \citealp{Kulkarni_2021} for an overview of past and future UV missions), there are none currently in operation. With this study we demonstrate that despite the lack of a currently operational, dedicated UV transient-searching facility, presently available UV instruments can be suitable for moderate-scale, systematic searches for UV transients. We show that such studies, combined with spectroscopic follow-up observations, may already lead to interesting discoveries in the field of UV transients.\

\begin{acknowledgements}
DM is partly supported by the Netherlands Research School for Astronomy (NOVA).\\

The SALT observations were obtained under the SALT Large Science Programme on transients (2018-2-LSP-001; PI: DAHB) which is also supported by Poland under grant no. MEiN 2021/WK/01. DAHB acknowledges research support from the National Research Foundation.\\

MG is supported by the EU Horizon 2020 research and innovation programme under grant agreement No 101004719.\\

We acknowledge the use of public data from the Swift data archive.\\ 

This research has made use of the VizieR catalogue access tool, CDS, Strasbourg, France (DOI : 10.26093/cds/vizier), and also of NASA's Astrophysics Data System Bibliographic Services.

\end{acknowledgements}

\bibliography{mybibliography.bib}

\end{document}